\documentclass[a4paper,10pt,final,color]{article}
 \usepackage{setspace}
\usepackage{geometry} 
\usepackage[utf8]{inputenc}
\usepackage[centertags]{amsmath}
\usepackage{amssymb}
\usepackage{amsthm}
\usepackage{amscd}
\usepackage{makeidx}
\usepackage{delarray}
\usepackage{amsmath}
\usepackage{colortbl}
\usepackage{anysize}
\usepackage{float}
\usepackage{graphicx}
\usepackage{enumerate}
\usepackage{latexsym,amstext,amscd}
\usepackage{fancyhdr}
\usepackage{amsfonts}
\usepackage{caption}
\usepackage{gensymb}

\usepackage{anysize}
\usepackage{capt-of}
\usepackage{multicol}
\usepackage{multirow}
\usepackage{sidecap}
\usepackage[font=small,skip=0pt]{caption}
\usepackage{subcaption}
\usepackage{cite}
\usepackage{pdfpages}
\usepackage{morefloats}
\usepackage[toc,page]{appendix}
\usepackage[labelfont=bf]{caption}
\usepackage{array}
\usepackage{emptypage}
\usepackage{lineno}

\usepackage{geometry}
\begin{document}

\begin{center}
 \textbf {\Large{Automatic quantification of abdominal subcutaneous and visceral adipose tissue in children, through MRI study, using total intensity maps and Convolutional Neural Networks}}
\vspace{0.5cm}

José Gerardo Suárez-García$^{1}$\footnote{The author JGSG was supported by the National Council of Sciences, Technologies and Humanities (CONAHCYT) to carry out this work, through a posdoctoral scholarship. }, Po-Wah So$^2$, Javier Miguel Hernández-López$^{1}$,  Silvia S. Hidalgo-Tobón$^{3,4}$, Pilar Dies-Suárez$^3$ and Benito de Celis-Alonso$^{1a}$
\vspace{0.3cm}

$^1$Facultad de Ciencias Físico-Matemáticas, Benemérita Universidad Autónoma de Puebla, Puebla, Mexico

$^2$Department of Neuroimaging, Institute of Psychiatry, Psychology and Neuroscience, King's College London, United Kingdom

$^3$Departamento de Imagenología, Hospital Infantil de México Federico Gómez, Mexico City, Mexico

$^4$Departamento de Física, Universidad Autónoma de México Iztapalapa, Mexico City, Mexico

$^a$bdca\_buap@yahoo.com.mx

\end{center}

\begin{abstract}

Childhood overweight and obesity is one of the main health problems in the world since it is related to the early appearance of different diseases, in addition to being a risk factor for later developing obesity in adulthood with its health and economic consequences.
Visceral abdominal tissue (VAT) is strongly related to the development of metabolic and cardiovascular diseases compared to abdominal subcutaneous adipose tissue (ASAT). Therefore, precise and automatic VAT and ASAT quantification methods would allow better diagnosis, monitoring and prevention of diseases caused by obesity at any stage of life. Currently, magnetic resonance imaging (MRI) is the standard for fat quantification, with Dixon sequences being the most useful. Different semiautomatic and automatic ASAT and VAT quantification methodologies have been proposed. In particular, the semi-automated quantification methodology used commercially through the cloud-based service AMRA\textsuperscript{\sffamily\textcircled{\tiny R}} Researcher (AMRA Medical AB, Linköping, Sweden) stands out due to its extensive validation in different studies. In the present work, a database made up of Dixon MRI sequences, obtained from children between 7 and 9 years of age, was studied. Applying a preprocessing to obtain what we call total intensity maps, a convolutional neural network (CNN) was proposed for the automatic quantification of ASAT and VAT. The quantifications obtained from the proposed methodology were compared with quantifications previously made through AMRA\textsuperscript{\sffamily\textcircled{\tiny R}} Researcher. For the comparison, correlation analysis, Bland-Altman graphs and non-parametric statistical tests were used. The results indicated a high correlation and similar precisions between the quantifications of this work and those of AMRA\textsuperscript{\sffamily\textcircled{\tiny R}} Researcher. The final objective is that the proposed methodology can serve as an accessible and free tool for the diagnosis, monitoring and prevention of diseases related to childhood obesity.
\end{abstract}

\section{Introduction}

Overweight and obesity in childhood is a global health problem. Between 2000 and 2016, the proportion of overweight children between the ages of 5 and 19 increased from 10\% to almost 20\%. Childhood overweight can lead to early onset of type 2 diabetes mellitus, as well as stigma and depression. In addition, childhood obesity is associated with an increased risk of obesity in adulthood, which has serious health and economic implications \cite{25}. Mexico is one of the main countries in the world with the highest pravelencia values. Using the body mass index (BMI) as reference, the prevalence of overweight in children between 5 and 9 years old (BMI $>$ 17.4) is equal to 18.8\%, while for obesity (BMI $>$ 19.8) it is equal to 18.6\% \cite{26}. However, for any BMI, each individual varies substantially in the distribution of body fat. This variation has important implications for the risk of developing different diseases \cite{1}. 

It is well known that higher amounts of visceral adipose tissue (VAT) compared to the amount of abdominal subcutaneous adipose tissue (ASAT) increase cardiovascular risk, development of type 2 diabetes mellitus, liver disease, cancer, and contracting infections (such as COVID -19) \cite{2}.
Quantitative, precise and reproducible measurements of total body fat and its distribution are therefore important for the prevention, diagnosis and monitoring of diseases related to overweight and obesity both in childhood and in adulthood \cite{3}.
 Dual-energy X-ray absorptiometry is a useful tool to accomplish this task. However, it makes modeling assumptions to differentiate VAT from ASAT, which produces errors in the quantifications. Also, it uses ionizing radiation and can only analyze 2D projections of the body \cite{3extra}. On the other hand, Magnetic Resonance Imaging (MRI) uses non-ionizing radiation and directly measuring total body fat content and distribution, as well as skeletal tissue mass accurately and reliably \cite{4}. Therefore, MRI is currently the gold standard for measuring body composition.
 In particular, the so-called Dixon technique is a rapid method that allows obtaining high-contrast images for soft tissue \cite{5}.
  This type of image uses the slight differences that exist between the magnetic resonance frequencies of the protons bound to the fat and water molecules, in order to distinguish the signals coming from each one. The set of images obtained from the Dixon sequences include in-phase, out-of-phase, fat-only, and water-only images from a single acquisition.
However, quantifying VAT and ASAT separately, both in children and adults, is still an interesting task to date, although the literature is still limited in studies of children \cite{5}.

Regarding semiautomatic analysis protocols, they have the disadvantage that they require the intervention of an operator or specialized personnel, resulting in a high cost, in addition to introducing variability depending on the analyst \cite{3}.
Different automatic VAT and SAT quantification methodologies have been proposed. Among them, those that apply Convolutional Neural Networks (CNNs) stand out both for the segmentation of the regions of interest, as well as for the quantification of fat deposits \cite{6,13,14}. CNNs are created specifically for image analysis. Its design aims to mimic the mechanism of the visual cortex of mammals, which is assumed to be formed by groups of ordered and specialized neurons for object recognition, starting from the simplest features to the most complex patterns \cite{7}  One of the advantages of CNNs is that they automatically learn the necessary image features, without the need to be entered by the user. CNNs have been applied to solve different problems such as the classification of brain tumors \cite{8}, detection of skin lesions \cite{9}, detection of diabetes through images of heart rhythm frequencies \cite{10}, breast cancer detection \cite{11}, COVID-19 detection through X-ray images \cite{12}, among many others.
 Recently, for example, Schneider et al. \cite{13}
proposed a software for automatic VAT and ASAT quantification and segmentation by studying MRI of adults applying UNet-based FCN architectures and data augmentation techniques, reaching high correlation values.
In another work, Devi et al. \cite{14} developed a hybrid convolutional neural network, combining a conventional CNN and a texture layer, for VAT and ASAT segmentation of abdominal MRI images of adults, obtaining a performance that, according to the authors, exceeds the state-of-the-art methods.  

Regarding studies in children, Armstrong et al. \cite{15} presented a paper in which they recognize that many conventional techniques applied in children to quantify body composition and liver fat have limitations, due to sensitivity to movement, mainly in the abdomen region due to breathing. Therefore, they developed a technique based on free-breathing radial and Cartesian MRI sequences to quantify body composition and hepatic proton-density fat fraction (PDFF) in children from 2 to 7 months of age, evaluating the feasibility for hepatic PDFF quantification using a scoring system made by a radiologist.
Also Armstrong et al. \cite{16}, in another study they compared non-sedated free breathing multi echo 3D stack of radial MRI versus standard breath holding and spectroscopy techniques for fat quantification. They studied healthy and overweight children between 7 and 13 years of age with nonalcoholic fatty liver disease, evaluating the quantifications using image quality scores, linear regression and Bland Altman analysis, obtaining accurate and repeatable measurements.
Kway et al. \cite{17} developed and evaluated an automatic segmentation method for the identification of abdominal adipose tissue (AAT), deep subcutaneous adipose tissue (DSAT) and visceral adipose tissue (VAT) deposits in neonates (less than two weeks old) and children. (ages between 4.5 and 6 years). Their method was based on a CNN based on the architecture known as U-net, which was compared with manual segmentations made by an expert through the calculation of Dice scores and Bland-Altman plots.

Among the semiautomatic quantification works, Peterli et al. \cite{18}, evaluated the distribution of visceral, subcutaneous, and liver fat in morbidly obese patients before and after bariatric surgery. In their work, they studied Dixon MRI sequences by applying automatic segmentation based on a statistical model (SSM), to later quantify ASAT, VAT and liver volumes through manual voxel counting. On the other hand, an outstanding semiautomatic methodology for quantifying fat and muscle compartments by studying DIXON sequences is the one used commercially through the cloud-based service AMRA\textsuperscript{\sffamily\textcircled{\tiny R}} Researcher (AMRA Medical AB, Linköping, Sweden). Its methodology has been described in detail and evaluated in terms of accuracy \cite{19,20,21,22,23}. This basically consists of the following. Image calibration to fat referenced images. Atlases with ground truth labels for fat and muscle compartments are recorded to an acquired MRI data set. Quality control is performed by trained operators, who can interactively adjust and improve the final segmentation. Finally, the volumes of fat and muscle are quantified within the segmented regions \cite{24}. Therefore, this methodology requires the intervention of an operator to perform quality control, and before performing the quantification, it is necessary to accurately segment the regions of interest. In addition, it is a commercial method, so an economic investment is necessary, making it not easily accessible to anyone.

 In the present work, a simple, economical and low computational methodology for the automatic quantification of VAT and ASAT was proposed. This was based on the study of Dixon sequences in phase, of male children between 7 and 9 years old, applying pre-processing techniques for the generation of what we call Total Intensity Maps. These maps included sufficient information on the regions of interest, and then, without the need to perform a precise segmentation, Convolutional Neural Networks (CNNs) proposed in two dimensions were applied to perform the quantifications. The reference standard were quantifications made previously through AMRA\textsuperscript{\sffamily\textcircled{\tiny R}} Researcher, comparing these with those obtained in this work, using Bland-Altmann plots, regression analysis and non-parametric statistical tests. 


\section{Methodology}

\subsection{Subjects}

In the present work, a proprietary database obtained from a collaborative project between researchers from institutions in Mexico and the United Kingdom was studied. This contained different MRI modalities of 78 mexican male children between 7 and 9 years of age, obtained at the Hospital Infantil de México in 2018. Among the children studied, 3 were underweight (BMI percentile $<$ 5), 42 normal weight (BMI percentile 5 - 85), 17 overweight (BMI percentile 85-95 ) and 16 obese (BMI percentile $>$95 ).
	
\subsection{MRI protocol}

All subjects were scanned using a Siemens 3T Skyra scanner (Syngo MR E11) (Siemens, Erlangen, Germany) with the dual-echo Dixon Vibe protocol, covering neck to knees. Subjects were scanned with five overlapping slabs of axial 3D spoiled gradient dual-echo images, in supine position with the arms along the sides and without localizer. Reconstruction of water-fat Dixon images was performed using the integrated scanner software. Common parameters for slabs one to three were: TR = 3.78 ms, TE = 1.23 ms, flip angle 10, bandwidth 123 Hz, 44 slices, voxel size 1.95$\times$1.95$\times$5 mm$^3$ and 256$\times$192 matrix, acquired during 17 seconds expiration breath-holds. Slabs four and five were acquired during free breathing with TR = 3.94 ms,  TE = 2.49 ms,  flip angle 10, bandwidth 123 Hz, 72 slices, voxel size 1.95$\times$1.95$\times$4 mm$^3$ and 256$\times$192 matrix. Viewed from the axial plane, each volume had dimensions of 192$\times$256$\times$44 voxeles.

\subsection{AMRA\textsuperscript{\sffamily\textcircled{\tiny R}} Researcher: semiautomatic quantification methodology}

For the 78 study subjects, body composition semiautomated quantification technique were performed from the reconstructed water and fat images using Dixon sequences in phase and out of phase, to later analyze these through the commercially available service AMRA\textsuperscript{\sffamily\textcircled{\tiny R}} Researcher. Briefly and as commented before, the analysis used in AMRA\textsuperscript{\sffamily\textcircled{\tiny R}} Researcher consisted of the following steps \cite{23}:
 (1)  Intensity inhomogeneity correction and calibration of fat and water images \cite{21}. (2)  Ground truth labels for fat compartments were registered to the acquired volumes  using non-rigid atlas based registration. (3)  All datasets were visually inspected and quality controlled by an trained analysis engineer at Advanced MR Analytics (Linkoping, Sweden), detecting and correcting common artifacts such as water-fat swaps (exchange of the signal-channel for fat and water due to ambiguities), anatomy outside field of view, breathing/motion artefacts, and issues with the MR-protocol.  (4) Quantification of fat, measured in liters (L), based on the calibrated images by integrating over the quality controlled labels. Finally, a report was generated. The included fat  compartments were visceral adipose tissue (VAT) and abdominal subcutaneous adipose
tissue (ASAT). VAT was defined as adipose tissue within the abdominal cavity, excluding adipose tissue outside the abdominal skeletal muscles and adipose tissue and lipids within and posterior of the spine and posterior of the back muscles. ASAT was defined as subcutaneous adipose tissue in the abdomen from the top of the femoral head to the top of the thoracic vertebrae T9. In each of the reports generated by AMRA\textsuperscript{\sffamily\textcircled{\tiny R}} Researcher, a precision (calculated from the coefficients of repeatability, i.e. the smallest detectable difference between two measurements at a 95\% confidence level) was declared equal to 0.17 L for VAT and equal to 0.33 L for ASAT.

\subsection{Proposed automatic quantification methodology}

In order to completely automate the quantification algorithm and avoid human intervention to correct the artifact known as water-fat swap, only in-phase Dixon sequences were studied. Remembering that each subject's scan was made up of five overlapping syllables, from top to bottom, only those numbered 2 and 3 were analyzed, since they contained the region of interest. Hereafter, these were called $V_1$ and $V_2$ respectively. Due to the overlap, the two volumes had to be joined by choosing the appropriate slice of each. Although the volumes were obtained in a single acquisition and with the indication of holding the breath, the joining process was not a trivial task. This was mainly due to artifacts caused by breathing or movement, causing differences between the range of intensities of both volumes, misalignment and mismatch in the anatomical regions. In order to correct this situation, a set of processes was proposed to correctly join the pair of volumes of each subject. All the algorithms presented in this work were developed with the MATLAB R2022b software, on a conventional computing system (Intel Core i7 12700H CPU, 16GB RAM, RTX 3070Ti GPU).

\subsubsection{Processes to join $V_1$ and $V_2$}

The intensities of the voxels of $V_1$ and $V_2$ were normalized, varying from 0 to 1 using the method known as min-max normalization. This was done assuming that between both volumes the voxels of lower intensity corresponded to the same type of tissue, as well as the voxels of higher intensity corresponded to another tissue. Considering the axial plane, the contrast of each volume was improved by histogram equalization. First, $V_2$ was completely equalized using as reference the histogram of the last 15 slices of $V_1$. Subsequently, $V_1$ was completely equalized using as reference the histogram of the first 15 slices of $V_2$ already equalized. Intensities less than 0.05 were set to 0, which corresponded to the empty bottom of each volume.

\begin{figure} 
	\begin{center} 
		\includegraphics[scale=0.4]{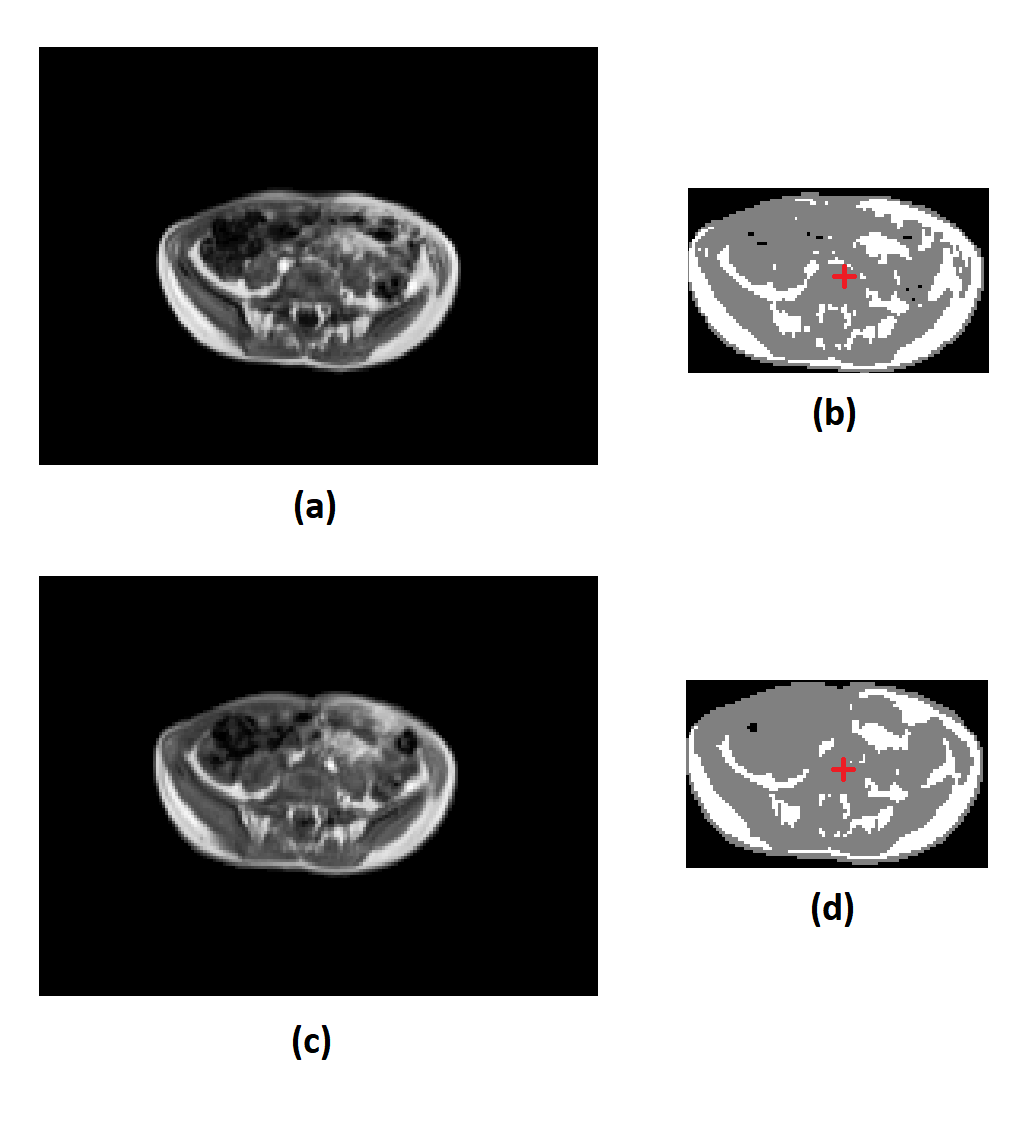} 
	\end{center} 
	\caption{{\bf Pair of slices chosen to join $V_1$ and $V_2$.} In (a) and (c) examples of slices of the volumes $V_1'$ and $V_2'$ (normalized and equalized) are shown respectively. In (b) and (d) the previous slices are shown with the voxels labeled within a box that completely contained them in volumes $V_1'$ and $V_2'$. These two slices were used to join the volumes since they obtained the highest value of the Dice coefficient. Furthermore, the slices served to center the volumes taking as reference the center of the chosen slices (red crosses).}
	\label{par_reb}
\end{figure}

\begin{figure} 
	\begin{center} 
		\includegraphics[scale=0.7]{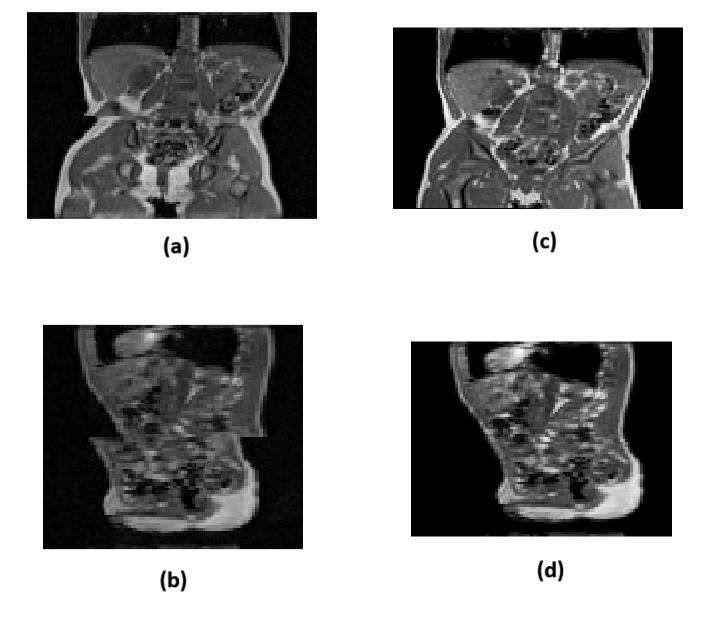} 
	\end{center} 
	\caption{{\bf Joining and centering of $V_1$ and $V_2$.} View from the coronal (a) and sagittal (b) planes of the join of $V_1$ and $V_2$ without applying any process to them. View from the coronal (c) and sagittal (d) planes of the normalized, equalized and centered $V_1$ and $V_2$, using as reference the pair of slices chosen from $V_1'$ and $V_2'$ respectively.}
	\label{ori_corr}
\end{figure}

In order to find the pair of slices (one from $V_1$ and another from $V_2$) that will serve to join the two volumes, only the last 8 slices of $V_1$ and the first 8 slices of $V_2$ were compared. These sets of slices were called $V_1'$ and $V_2'$, which had dimensions of 192$\times$256$\times$8 voxels. 
In each volume different regions of the body that were not of interest could be visible, such as arms, shoulders and hands. So, before comparing volumes $V_1'$ and $V_2'$, it was necessary to apply an algorithm to exclude the mentioned regions. Considering that these regions appeared separated from the region of interest, the algorithm simply started from a voxel located approximately in the central area contained in the region of interest and only all neighboring voxels that were connected to each other were retained. Once this was done, the comparison between $V_1'$ and $V_2'$ continued. A box with the smallest dimensions was sought such that it completely contained either of the two volumes $V_1'$ and $V_2'$. Voxels with intensities greater than 0 were labeled using a threshold equal to 0.5, such that voxels with intensities less than or equal to 0.5 were labeled as 1, and voxels with intensities greater than 0.5 were labeled as 2. Next, the 8 slices of $V_1'$ and the 8 slices of $V_2'$, already labeled, were compared in pairs by calculating the so-called Dice coefficient. From the 64 comparisons made, the pair of slices that obtained the highest value of the Dice coefficient were used as a reference to join the two complete volumes $V_1$ and $V_2$ (Fig. \ref{par_reb}). In addition to joining the volumes, they were also centered. To do this, using the chosen slices, the pair of voxels located in the center of them were used as reference points to center and finally join the volumes $V_1$ and $V_2$ (Fig. \ref{ori_corr}). After centering and joining $V_1$ and $V_2$, both volumes ended up displaced relative to each other. However, the joined volume had to be contained in a single volume with uniform dimensions. To do this, the two slices that served to join $V_1$ and $V_2$ were centered within two slices of dimensions 200x200 voxels respectively. Then, these slices were joined, and subsequently the rest of the volumes were contained in a single volume with dimensions in the axial plane of 200x200 voxels and height equal to the sum of the heights of the two joined volumes.

\begin{figure} 
	\begin{center} 
		\includegraphics[scale=0.33]{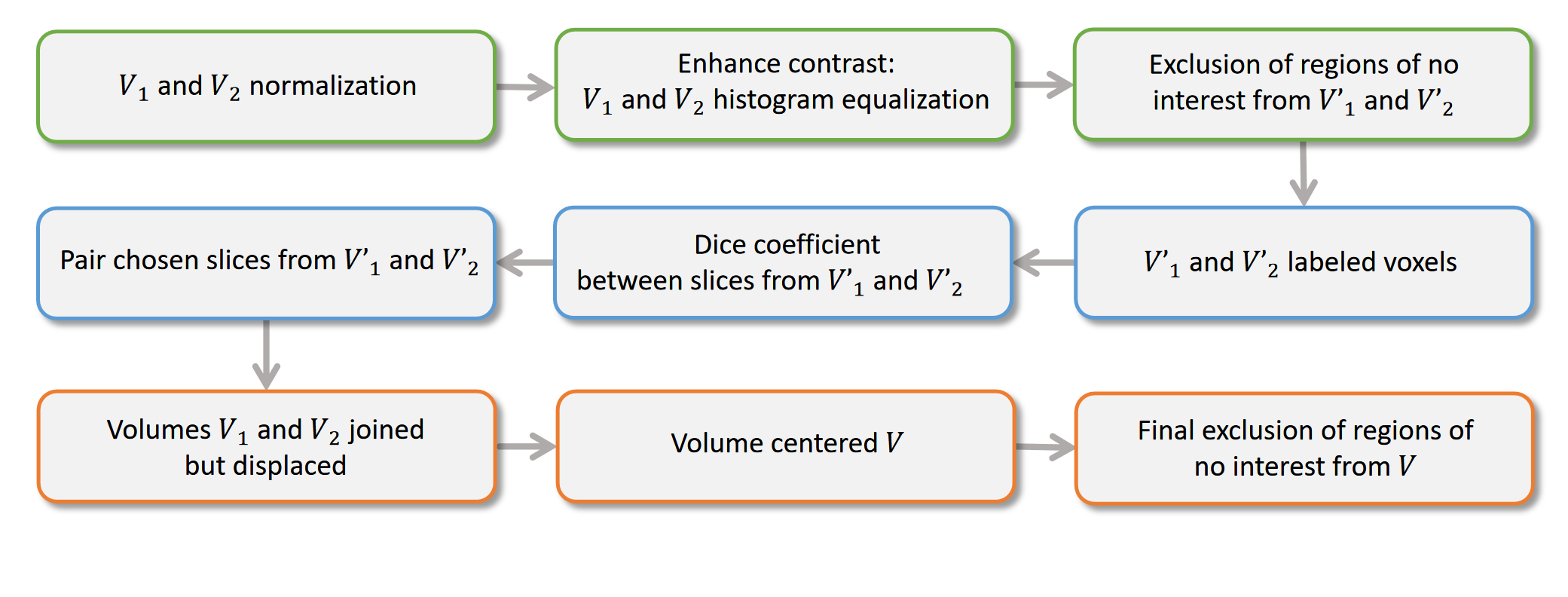} 
	\end{center} 
	\caption{{\bf Volume joining.} Processes carried out to find the pair of slices from $V_1'$ and $V_2'$ that served as a reference to join the two volumes $V_1$ and $V_2$ and thus obtain, for each subject, a single joined volume $V$ normalized, equalized and centered.}
	\label{diag_par}
\end{figure}

Afterwards, only 30 total slices of the joined volume were retained, with 10 from $V_1$ starting from its chosen slice upward, and 20 from $V_2$ starting from its chosen slice downward. The joined volume was called $V$, which had dimensions of 200$\times$200$\times$30 voxels. As mentioned before, regions that were not of interest were excluded from volumes $V_1'$ and $V_2'$ (with 8 slices each). Then, with the joined volume $V$ (having 30 slices in total), this task was repeated in a slightly different way. Instead of choosing a voxel located in the center of the entire volume, voxels located in the center of each of the 30 slices were searched. In each slice separately, starting from the central voxel, only the voxels that were connected to it and to each other were added. Because the volume was already centered, performing this task for each slice was more efficient than performing it at once for the entire volume. Fig. \ref{diag_par} shows a diagram with all the processes followed to join the volumes.

\subsubsection{Creation of total intensity maps $I_{asat}$ and $I_{vat}$ for training the proposed CNNs}

From each joined volume $V$, two-dimensional maps $I_{asat}$ and $I_{vat}$ were created. These two together formed a new volume $V_{I}$ with dimensions 200$\times$200$\times $2 voxels. The volumes $V_{I}$ were used as inputs to two proposed two-dimensional CNNs whose tasks were the quantification of ASAT and VAT respectively. The volumes $V_{I}$ were considered by the CNNs as 2D images with two different channels. The image $I_{asat}$ was created from a volume $V_{asat}$, which contained an approximate segmentation of the region where the ASAT should be located. On the other hand, the image $I_{vat}$ was created from a volume $V_{vat}$, which contained an approximate segmentation of the region where the VAT should have been located. The images $I_{asat}$ and $I_{vat}$ were called total intensity maps. To obtain these two images, the following was done for each subject.

\begin{figure} 
	\begin{center} 
		\includegraphics[scale=0.31]{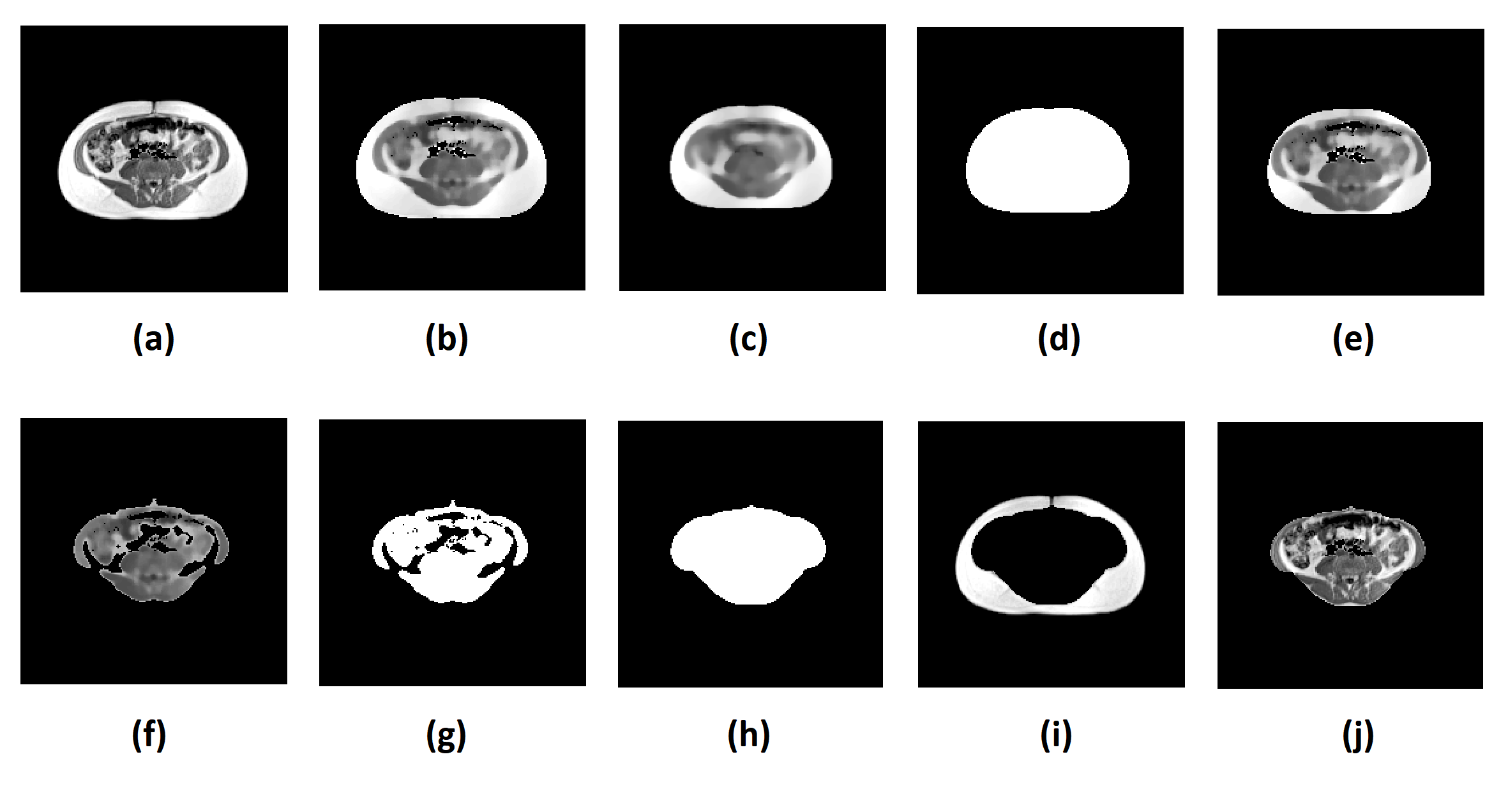} 
	\end{center} 
	\caption{{\bf Approximate segmentation of the regions that should have contained the VAT and ASAT.} (a) Slice of a volume $V$. (b) Smoothing. (c) Volume resized and smoothed again. (d) Voxels set to 1, application of hole filling process and mask creation. (e) Volume obtained after applying the mask to the volume shown in (b), eliminating voxels located outside it. (f) Elimination of voxels with intensities greater than 0.75. (g) Voxels equal to 1. (h) Application of hole filling processes and creation of new mask. Applying this last mask to the volume shown in (a), the volume (i) $V_{asat}$ was obtained by eliminating the voxels inside it, and the volume (j) $V_{vat}$ by eliminating the voxels outside it. }
	\label{seg_vat_asat}
\end{figure}

The volume $V$ was smoothed with a median filter of size 3$\times$3$\times$3 voxels and was subsequently normalized from 0 to 1. Then, a resized volume proportional to 85\% of the original volume was obtained. This volume was smoothed with a median filter of size 7$\times$7$\times$7 voxels. All voxels with intensities greater than 0 were set equal to 1. A filling process was carried out to eliminate possible holes in the resized volume. This volume was used as a mask over the original volume $V$, so that all voxels within the mask with intensities greater than a threshold equal to 0.75 were set to 0. The remaining voxels were set to 1, and again a filling process to eliminate possible holes was applied. This volume served as a new mask over the original volume $V$ used in the following way. By eliminating the voxels that were inside the mask, the volume $V_{asat}$ was obtained, which contained an approximate segmentation of the region that must have contained the ASAT. On the other hand, by conserving only the voxels that were within the last mask, the volume $V_{vat}$ was obtained, which contained an approximate segmentation of the region that should have contained the VAT. An example of the process described above is shown in Fig. \ref{seg_vat_asat}. Although the figure shows a slice as an example, the process was performed with the entire volume $V$ at the same time.

\begin{figure} 
	\begin{center} 
		\includegraphics[scale=0.45]{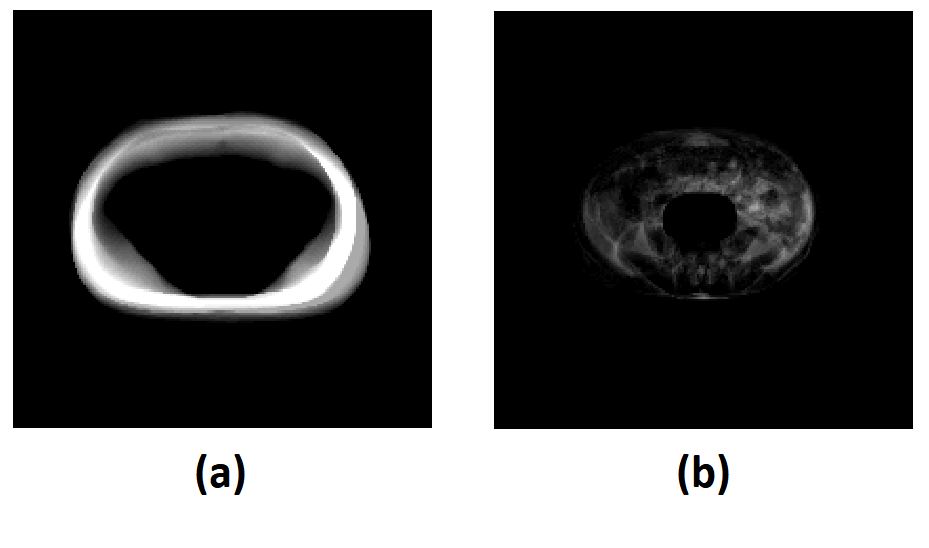} 
	\end{center} 
	\caption{{\bf  Total intensity maps.} (a) Total intensity map $I_{asat}$, obtained from the volume $V_{asat}$. (b) Total intensity map $I_{vat}$, obtained from the volume $V_{vat}$. Both maps formed the slices of a volume $V_{I}$ of dimensions 200$\times$200$\times$2 voxels that was used as input to the proposed CNNs.}
	\label{im_suma}
\end{figure}

To obtain the total intensity maps $I_{asat}$ and $I_{vat}$ the following was done. All voxels of the volume $V_{asat}$ whose intensities were different from 0 were set equal to 1 and the following equation was applied:

\begin{equation}
I_{asat}(x,y) = \sum^{30}_{z = 1}V_{asat}(x,y,z) 
\end{equation}

\noindent so that all the intensities of the voxels located in the same position of each slice with dimensions equal to 200$\times$200 voxels were added. Thus, the total intensity map $I_{asat}$ with dimensions equal to 200$\times$200$\times$1 voxels was obtained (Fig. \ref{im_suma}(a)). To obtain the second map, the following was done. On the other hand, all voxels of the volume $V_{vat}$ whose intensities were less than 0.7 were set to 0, while voxels with intensities greater than or equal to 0.7 retained their value. Then, the following equation was applied:

\begin{equation}
I_{vat}(x,y) = \sum^{30}_{z = 1}V_{vat}(x,y,z)
\end{equation}

\noindent so that all the intensities of the voxels located in the same position of each slice with dimensions equal to 200$\times$200 voxels were also added. Thus, the total intensity map $I_{vat}$ was obtained with dimensions equal to 200$\times$200$\times$1 voxels. Finally, a volume $V_{I}$ was created from the two aforementioned maps. This volume had $I_{asat}$ as its first slice and $I_{vat}$ as its second, thus forming a volume with dimensions 200$\times$200$\times$2 voxels (Fig. \ref{im_suma}(b )). The volumes $V_{I}$ obtained from each subject were used as inputs for the proposed CNNs that will be described in the following section.

\subsubsection{Proposed CNNs}

Two CNN architectures were proposed to quantify ASAT and VAT respectively. Both CNNs had a similar structure and studied the same volumes $V_{I}$ of each subject. From 78 subjects, 42 were randomly chosen for training, 18 for validation and 18 for testing. Table \ref{suj_imc} shows their distribution according to their weight classification.

\begin{table}
	\centering
	
	\begin{tabular}{cccccc}
		\hline 
	\textbf{Subset}  & \textbf{Low weight}  &	  \textbf{Normal weight}   &	  \textbf{Overweight}  &	  \textbf{Obesity} &	  \textbf{Total} \\				
		\hline  
		Training & 3 & 20 & 8 & 11 & 42 \\
         Validation & 0 & 11 & 5 & 2 & 18\\
          Testing & 0 & 11 & 4 & 3 & 18\\
Total & 3 & 42 & 17 & 16 & 78 \\
		\hline
	\end{tabular}
	\caption{{\bf Distribution of subjects according to their weight classification.}}
	\label{suj_imc}
\end{table}

\begin{figure} 
	\begin{center} 
		\includegraphics[scale=0.22]{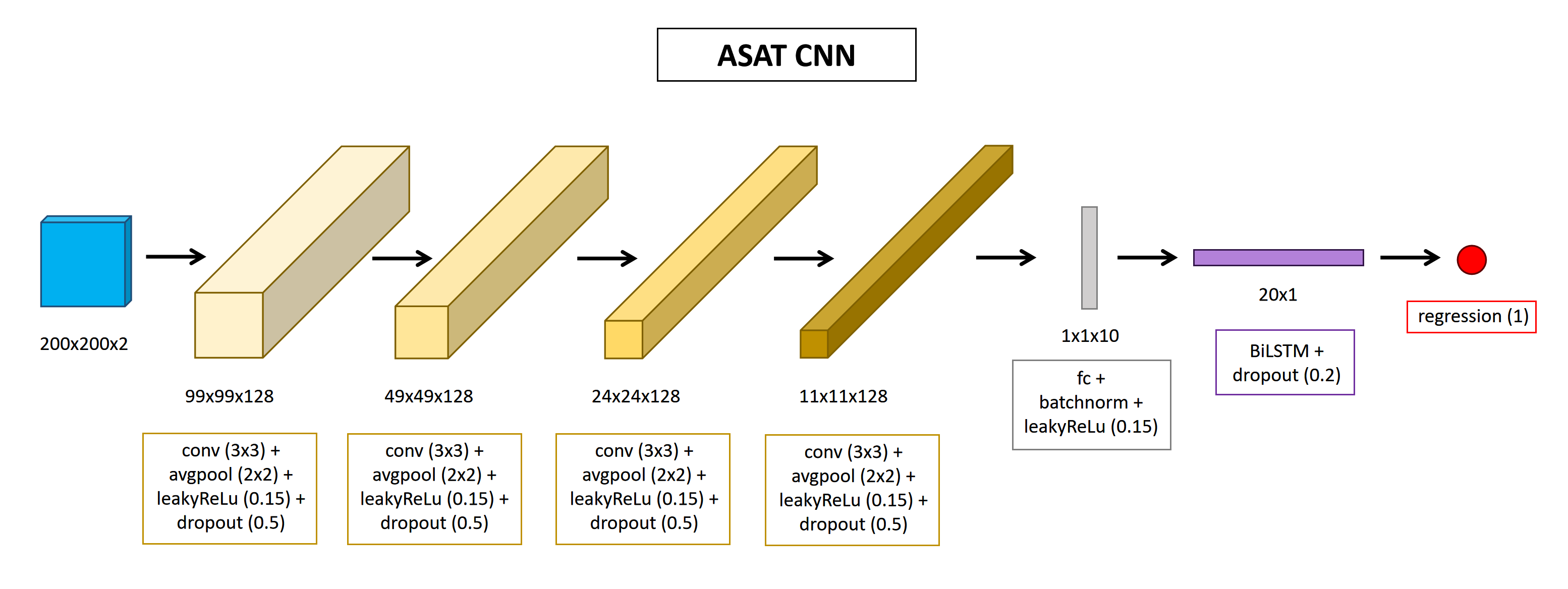} 
	\end{center} 
	\caption{{\bf CNN architecture to quantify the ASAT.} The different blocks and layers that formed the proposed CNN are shown.}
	\label{cnn_asat}
\end{figure}

Fig. \ref{cnn_asat} shows the architecture of the CNN to quantify the ASAT. There were four blocks with the same layers. The first was a convolution layer with 128 filters of size 3$\times$3 with stride equal to 1$\times$1; the second was an average pooling layer of size 2$\times$2 with stride equal to 2$\times$2 and same padding; the third was a Leaky ReLu activation layer with scale equal to 0.15; and the fourth was a dropout layer with probability equal to 0.5. After the mentioned blocks, there was a fully connected layer of 10 nodes, followed by a batch normalization layer and a Leaky ReLu activation layer with scale equal to 0.15. Then there was a BiLSTM layer with 10 hidden units and 20$\times$1 hidden states and a dropout layer with probability 0.2. Finally there was a regression layer with a single output node. To avoid overfitting, data augmentation was used through random rotations varying from -30 to 30 degrees. A minibatch equal to 42 was used (this being the total number of training samples), applying the SGDM optimizer with a constant learning range equal to 0.005. The CNN was trained for 30,000 epochs with validations every 10 epochs.

\begin{figure} 
	\begin{center} 
		\includegraphics[scale=0.22]{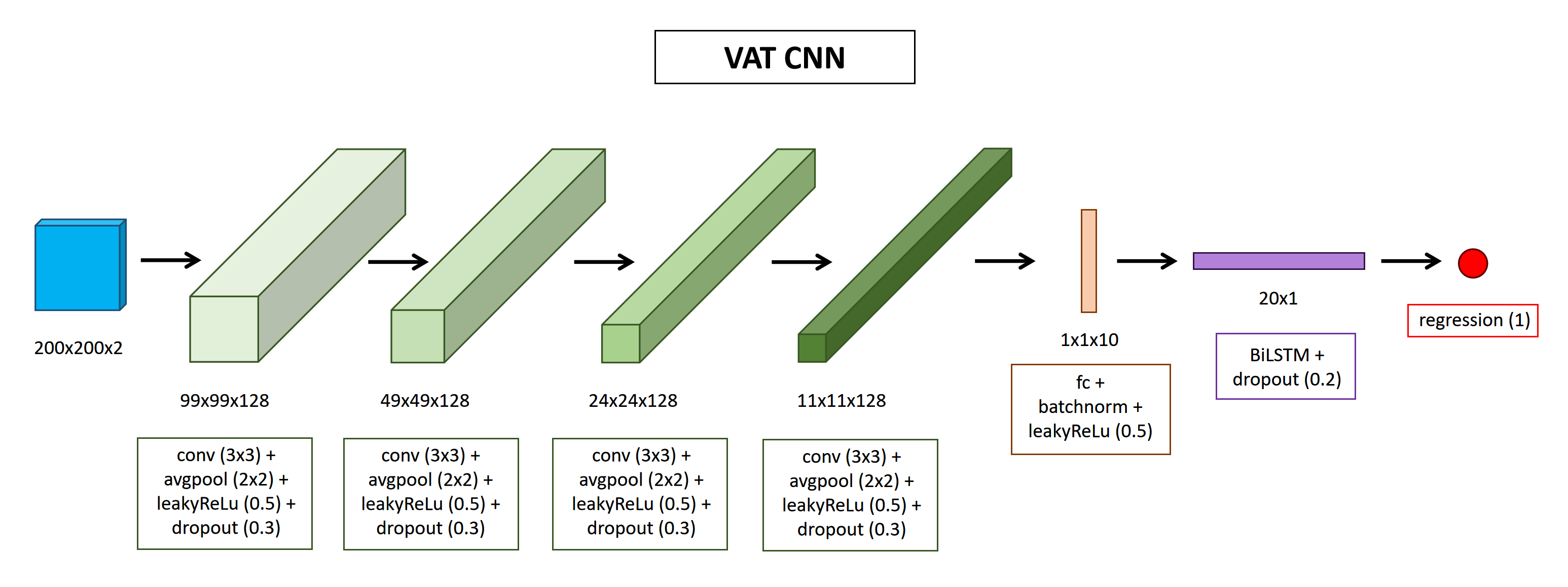} 
	\end{center} 
	\caption{{\bf CNN architecture to quantify the VAT.} The different blocks and layers that formed the proposed CNN are shown.}
	\label{cnn_avat}
\end{figure}

Fig. \ref{cnn_avat} shows the CNN architecture to quantify the VAT. Its architecture was similar to that for quantifying the ASAT. Their differences were the following. For the second CNN, Leaky ReLu activation layers with scale equal to 0.5 were used; the probability of dropout layers were equal to 0.3; rotation angles for augmentation ranged from -10 to 10 degrees; and the constant learning range was equal to 0.001. The expected outputs of each CNN were the quantizations reported by AMRA\textsuperscript{\sffamily\textcircled{\tiny R}} Researcher. After training the CNNs, they were applied to the training, validation and testing subjects. In order to compare the quantifications between AMRA\textsuperscript{\sffamily\textcircled{\tiny R}} Researcher and those made by the CNNs, Bland-Altman plots were created, correlation analysis was performed and the non-parametric statistical test called Wilcoxon signed rank test was applied.

\section{Results}

\begin{figure} 
	\begin{center} 
		\includegraphics[scale=0.365]{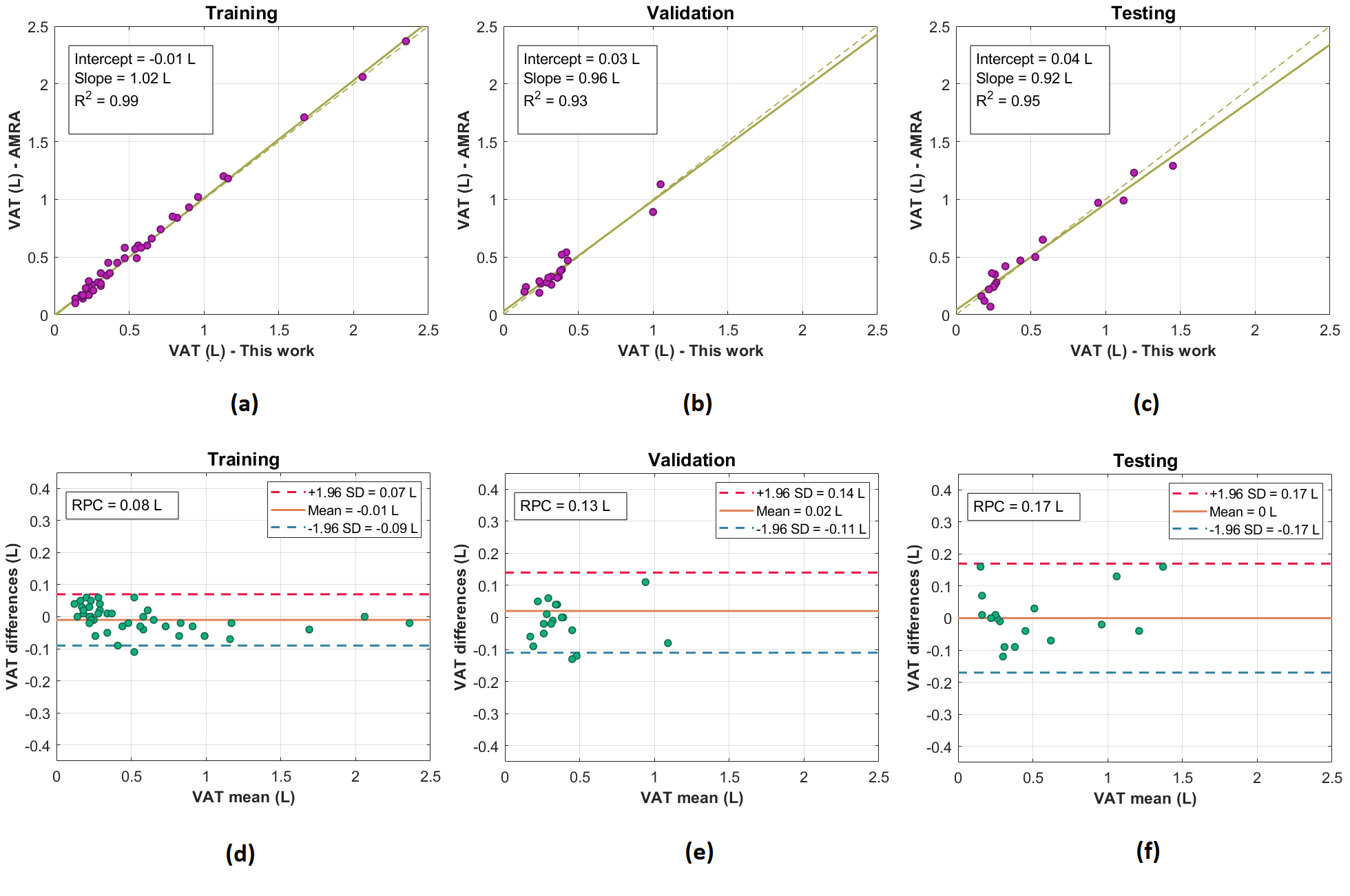} 
	\end{center} 
	\caption{{\bf Bland-Altman and correlation plots for VAT.} (a), (b) and (c) show the correlation plots, while (d), (e) and (f) show the Blan-Altman plots, for the training, validation and testing subjects respectively for the quantification of VAT.}
	\label{bland_vat}
\end{figure}

\begin{figure} 
	\begin{center} 
		\includegraphics[scale=0.36]{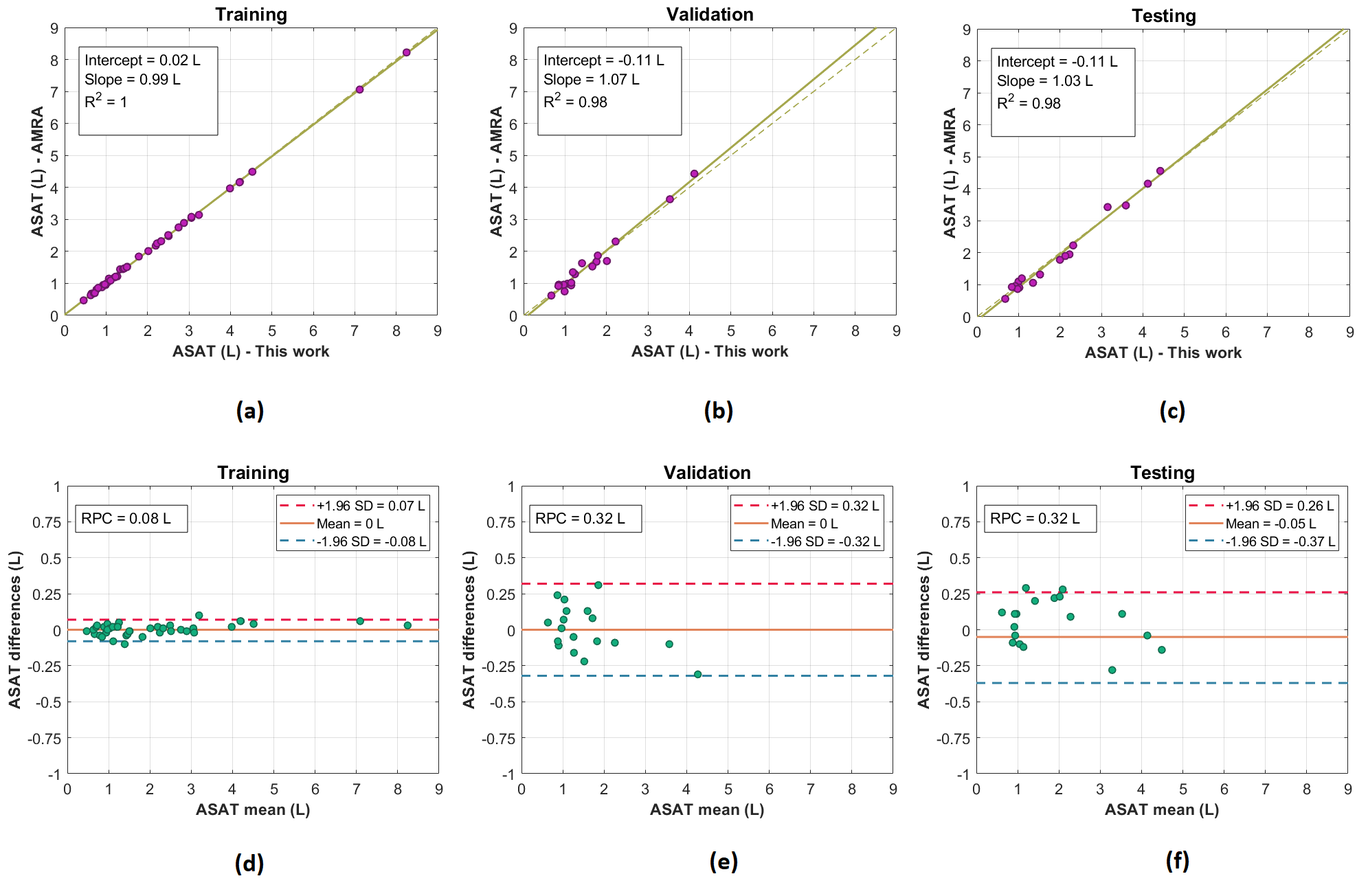} 
	\end{center} 
	\caption{{\bf Bland-Altman and correlation plots for ASAT.} (a), (b) and (c) show the correlation plots, while (d), (e) and (f) show the Blan-Altman plots, for the training, validation and testing subjects respectively for the quantification of ASAT.}
	\label{bland_asat}
\end{figure}

Fig. \ref{bland_vat} shows the correlation and Bland-Altman plots obtained by comparing the VAT quantifications made by AMRA\textsuperscript{\sffamily\textcircled{\tiny R}} Researcher and those obtained in the present work, this for the training, validation and testing subjects respectively. Something similar is shown in Fig. \ref{bland_asat} when comparing the ASAT quantifications. The correlation graphs indicated the value of $R^2$, the intercept and slope of the fit line.  
Table \ref{pcorr} shows the p-values that indicated whether there was no significant correlation between the compared quantifications (null hypothesis).
All results indicated a high correlation which was statistically significant. Bland-Altman plots indicated the average difference of the quantifications, the 95\% limits of agreement and the reproducibility coefficient (RCP). For the quantification of VAT, coefficients of variation (CV) equal to 7.3\%, 16\% and 17\%, and RCP equal to 0.08 L, 0.13 L and 0.17 L, were obtained for the training, validation and testing subjects. respectively. For the quantification of ASAT, CV equal to 1.8\%, 10\% and 8.6\%, and RCP equal to 0.08 L, 0.32 L and 0.32 L, were obtained for the training, validation and testing subjects respectively. Table \ref{wilcoxon} shows the results from the Wilcoxon signed rank test between the AMRA\textsuperscript{\sffamily\textcircled{\tiny R}} Researcher quantifications and those made in the present work, for the VAT and ASAT, with the training, validation and test subjects respectively. All tests indicated that there were no significant statistical differences (p-value $>$ 0.05).


\begin{table}
	\centering
	
	\begin{tabular}{ccc}
		\hline 
	\textbf{Subset}  & \textbf{p-value (VAT)}  &	  \textbf{p-value (ASAT)}    \\				
		\hline  
		Training & 0 & 0  \\
         Validation & 9.503$\times$10$^{-11}$ & 1.511$\times$10$^{-14}$ \\
          Testing & 3.586$\times$10$^{-12}$ & 1.039$\times$10$^{-15}$\\

		\hline
	\end{tabular}
	\caption{{\bf Correlation p-values.} p-values obtained to know if the correlations between AMRA\textsuperscript{\sffamily\textcircled{\tiny R}} Researcher quantifications and those made in the present work were significant.}
	\label{pcorr}
\end{table}

\begin{table}
	\centering
	
	\begin{tabular}{ccc}
		\hline 
	\textbf{Subset}  & \textbf{p-value (VAT)}  &	  \textbf{p-value (ASAT)}    \\				
		\hline  
		Training & 0.2998 & 0.3093  \\
         Validation & 0.2656 & 0.9478 \\
          Testing & 0.7757 & 0.1765\\

		\hline
	\end{tabular}
	\caption{{\bf Results from the Wilcoxon signed rank test.} The p-values obtained after applying the Wilcoxon signed rank test between the AMRA\textsuperscript{\sffamily\textcircled{\tiny R}} Researcher quantifications and those made in the present work are shown.}
	\label{wilcoxon}
\end{table}

\section{Discussion and conclutions}

The present work used as a reference standard the quantifications made by the widely validated commercial measurement system called AMRA\textsuperscript{\sffamily\textcircled{\tiny R}} Researcher. Within the AMRA\textsuperscript{\sffamily\textcircled{\tiny R}} Researcher reports (obtained during 2018, the year on which the measurements were carried out), precisions equal to 0.17 L and 0.33 L were indicated for the quantifications of VAT and ASAT respectively. This precision was defined as a repeatability coefficient, that is, the smallest detectable difference between two measurements with a confidence level of 95\%, made under the same conditions, with the same imaging protocol and quantification methodology. In this work, the same Dixon sequences used by AMRA\textsuperscript{\sffamily\textcircled{\tiny R}} Researcher were studied (so the same imaging protocol was used). However, a different quantification methodology was proposed. Therefore, to make a comparison between the AMRA\textsuperscript{\sffamily\textcircled{\tiny R}} Researcher quantifications and those made in this work, the reproducibility coefficient (RPC) was used, which is defined as the value under which the absolute differences between two measurements would fall within 95 \% probability, considering that these were calculated under different conditions or using different measurement systems \cite{27}.

As shown in the results, for the quantification of VAT a RCP $\leq 0.17$ L was obtained, while for the quantification of ASAT a RCP $\leq 0.32$ L was obtained. Therefore, it can be concluded that the Measurements made in this work were within the precision reported by AMRA\textsuperscript{\sffamily\textcircled{\tiny R}} Researcher. Although today the precision of AMRA\textsuperscript{\sffamily\textcircled{\tiny R}} Researcher may possibly be greater, it would be necessary to obtain new quantifications carried out through its updated methodology, compare them with those made in the present work and then compare the precision of both methodologies.

Although the work depended on DIXON sequences generated using the AMRA\textsuperscript{\sffamily\textcircled{\tiny R}} Researcher imaging protocol, the proposed methodology could be verified on other databases. This is because these types of sequences are commonly obtained by different resonators. However, there may be an exception regarding the way slabs are obtained, since this procedure is specific to AMRA\textsuperscript{\sffamily\textcircled{\tiny R}} Researcher. Even considering the above, in general it would be more convenient to study a single volume obtained at once containing the entire region of interest, omitting the process of joining slabs, and reducing errors in quantifications due to errors made during the joining process. In this case, the proposed methodology could be applied, excluding the union process and appropriately choosing the 30 slices from which the total intensity maps would be obtained.

The proposed two-dimensional CNNs analyzed a set of volumes $V_{I}$ formed by what we call $I_{vat}$ and $I_{asat}$. These maps were obtained by adding the intensities of the voxels that resulted from approximately segmenting the regions that should have contained the VAT and ASAT respectively. Then, the proposed CNNs considered the volumes $V_{I}$ as a 2D image with two different channels. Training CNNs in two dimensions required a much smaller amount of computational resources compared to studying 3D volumes. Therefore, this was an advantage of the proposed two-dimensional methodology. Furthermore, the proposed CNNs had much simpler architectures than many others used by different works, obtaining excellent results in this case. On the other hand, after studying the training, validation and testing subjects, it was observed that there was consistency in the results, thus demonstrating their reproducibility.  When the Dixon-in-phase sequences were studied, the voxels with high intensities did not correspond solely to the fat signal, since it could have been a combination of this with the water signal. Furthermore, volumes $V_{I}$ were made up of a small number of slices (30 in total), so they did not necessarily cover the entire region that contained the VAT and ASAT. Also, no anatomical reference was used, except for the choice of slabs numbered 2 and 3. Methodologies from other works (including AMRA\textsuperscript{\sffamily\textcircled{\tiny R}} Researcher) needed to first accurately segment the fat deposits, then decided which ones were part of VAT and ASAT, and finally quantified them. For the quantifications of this work, it was hypothesized that the approximate segmentation made was sufficient to implicitly relate it to the amount of fat to be studied. The proposed methodology had only the objective of quantifying fat without having to locate or segment it precisely. It was known that the main strength of CNNs was the automatic search for abstract patterns in images, with the aim of successfully performing various tasks such as segmentation, classification or detection. Therefore, through the adequate training of the proposed CNNs using the total intensity maps $I_{asat}$ $I_{vat}$ as inputs, without requiring precise segmentations and without further anatomical considerations, it was possible to successfully quantify the VAT and ASAT with similar accuracy to AMRA\textsuperscript{\sffamily\textcircled{\tiny R}} Researcher.


Among the limitations of this work, it is found that the database studied was made up of a small number of samples. Additionally, the study subjects had different BMIs, and there was an imbalance between the total number of samples for each weight classification. Although accurate results were obtained, it can be deduced that the methodology had a bias towards subjects with a normal weight, since those were the ones who had the greatest number of samples during training. On the other hand, this work analyzed the DIXON sequence in phase, thus avoiding the necessary correction of the artifact known as water-fat swap, but wasting the use of fat-only images which contained more useful and explicit information to perform the quantifications. Also, the total intensity maps $I_{asat}$ and $I_{vat}$ lost spatial information when reducing a 3D volumes to a 2D images. Therefore, these maps were mostly affected by artifacts generated by movement. As this was a study conducted in children between 7 and 9 years old, the appearance of these artifacts was more likely since the breath-hold condition may not always have been met, nor the subjects were completly at rest during the complete acquisition of the MRI sequences.

Future work should apply the proposed methodology in databases with a greater number of samples with balanced classes and performing cross-validations. Since the study was restricted to male children aged between 7 and 9 years, the proposed method could be applied to subjects of different ages, both children and adults, as well as men and women. Furthermore, fat-only DIXON sequences should be studied, proposing an automatic method for correcting the water-fat-swap artifact and thus taking advantage of the information that this type of sequence offers for the required quantification tasks. Also, an algorithm could be implemented which would automatically choose the best anatomical region of the volumes to perform the quantification, so that the total intensity maps $I_{asat}$ and $I_{vat}$ would contain a greater amount of useful information for train the CNNs. Finally, other CNN architectures could be proposed.

In conclusion, an automatic, simple, reproducible and economical methodology for quantifying ASAT and VAT in children was proposed, with low demand for computational resources, based on the analysis of what we called total intensity maps and CNNs in two dimensions with a simple architecture, achieving the precision of the commercial AMRA\textsuperscript{\sffamily\textcircled{\tiny R}} Researcher quantification method. In this work, Dixon sequences commonly obtained in different scanners were studied, making the proposed methodology accessible and reproducible by independent studies, in order to corroborate the results and implement improvements. In the end, all of the above had the final objective that the proposed methodology can serve as an accessible and free tool for the diagnosis, monitoring and prevention of diseases related to overweight and obesity in children.

\end{document}